\documentstyle[emulateapj]{article}

\def\lae{\mathrel{<\kern-1.0em\lower0.9ex\hbox{$\sim$}}}
\def\gae{\mathrel{>\kern-1.0em\lower0.9ex\hbox{$\sim$}}}

\def\tg{\phantom{1}}
 
\slugcomment{Accepted for publication in the Astronomical Journal, August 1999 issue.}
 
\lefthead{C\^ot\'e}
\righthead{Kinematics of the Galactic Globular Cluster System}
 
\begin{document}
 
\title{Kinematics of the Galactic Globular Cluster System: New 
Radial Velocities for Clusters in the Direction of the Inner Galaxy}
 
\author{Patrick C\^ot\'e\altaffilmark{1}}
\affil{California Institute of Technology, Mail Stop 105-24, Pasadena, CA 91125\\
     {\rm pc@astro.caltech.edu}}

\altaffiltext{1}{Sherman M. Fairchild Fellow}

% Notice that each of these authors has alternate affiliations, which
% are identified by the \altaffilmark after each name.  The actual alternate
% affiliation information is typeset in footnotes at the bottom of the
% first page, and the text itself is specified in \altaffiltext commands.
% There is a separate \altaffiltext for each alternate affiliation
% indicated above.
 
% The abstract environment prints out the receipt and acceptance dates
% if they are relevant for the journal style.  For the aasms style, they
% will print out as horizontal rules for the editorial staff to type
% on, so long as the author does not include \received and \accepted
% commands.  This should not be done, since \received and \accepted dates
% are not known to the author.
 
\begin{abstract}
The High Resolution Echelle Spectrometer (HIRES) on the Keck I telescope has been used to 
measure the first radial velocities for stars belonging to eleven, heavily-reddened
globular clusters in the direction of the inner Galaxy.  The sample consists of the 
clusters Terzan 3, NGC 6256, IC 1257, NGC 6380 (= Ton 1), Ton 2 (= Pismis 26), Djorg 1, 
NGC 6540 (= Djorg 3), IC 1276 (= Pal 7), Terzan 12, NGC 6749 and Pal 10.
Candidate cluster members were selected from a combination of previously published
color-magnitude diagrams (CMDs) and new instrumental CMDs obtained with the
Palomar 1.5m telescope.
The systemic velocities of Djorg 1 and Pal 10 should be considered provisional since 
velocities are available for only two stars. For the remaining nine clusters, we have 
measured radial velocities for three to nine member stars.
Using our HIRES spectra, we estimate metallicities of [Fe/H] $\simeq$
$-0.75$ for both Terzan 3 and IC 1276: two cluster lacking previous metallicity
estimates.  The question of kinematic substructuring among the Galactic globular cluster 
system is investigated using an updated catalog of globular cluster distances, 
metallicities and velocities.  It is found that the population of metal-rich globular 
clusters shows significant rotation at all Galactocentric radii. For the metal-rich clusters 
within 4 kpc of the Galactic center, the measured rotation velocity and line-of-sight 
velocity dispersion are similar to those of bulge field stars.
We investigate claims that the metal-rich 
clusters are associated with the central Galactic bar by comparing the kinematics of
the innermost clusters to that of the atomic hydrogen in the inner Galaxy. 
The longitude-velocity diagram of both metal-rich and metal-poor clusters bears a remarkable 
similarity to that of the gas, including the same non-circular
motions which have traditionally been interpreted as evidence for a Galactic bar,
or, alternatively, a non-axisymmetric bulge. However, uncertainties in the 
existing three-dimensional Galactocentric positions for most of the clusters 
do not yet allow an unambiguous discrimination between the competing
scenarios of membership in a rigidly rotating bar, or in a bulge which is an oblate isotropic rotator.
We conclude that the majority of metal-rich clusters within the central $\sim$ 4 kpc 
of the Galaxy are probably associated with the bulge/bar, and not the thick disk.
\end{abstract}
 
% The different journals have different requirements for keywords.  The
% keywords.apj file, found on aas.org in the pubs/aastex-misc directory, 
% contains a list of keywords used with the ApJ and Letters.  These are 
% usually assigned by the editor, but authors may include them in their 
% manuscripts if they wish. 
 
 
% That's it for the front matter.  On to the main body of the paper.
% We'll only put in tutorial remarks at the beginning of each section
% so you can see entire sections together.
 
% In the first two sections, you should notice the use of the LaTeX \cite
% command to identify citations.  The citations are tied to the
% reference list via symbolic KEYs.  We have chosen the first three
% characters of the first author's name plus the last two numeral of the
% year of publication.  The corresponding reference has a \bibitem
% command in the reference list below.
%
% Please see the AASTeX manual for a more complete discussion on how to make
% \cite-\bibitem work for you.   

\section{Introduction}

The spatial distribution, chemical abundance, age and kinematics of Galactic
star clusters provide some of the most powerful constraints on models
for the formation and evolution of the Milky Way. As the oldest
and most metal-poor of these objects, globular clusters (GCs) have received
particularly close scrutiny since their properties hold important clues to the
structure and evolution of the Galaxy and its various components 
at the earliest possible epochs.

Most of the GCs within $\sim 25^{\circ}$ of the Galactic center have high metallicities 
relative to their counterparts at large Galactocentric distances.
Although it has been recognized for some time that the Galactic GC system shows a bimodal
distribution in metallicity (Morgan 1959; Kinman 1959), and that the metal-poor component is
clearly associated with the Galactic halo, the proper categorization
of the metal-rich GCs is less certain. Early suggestions that the metal-rich 
($i.e.$, G-type) GCs formed a flattened, disk-like system ($e.g.$, Baade 1958; Morgan 1959;
Kinman 1959) were called into question by Woltjer (1975), Harris (1976) and Frenk \& White
(1982) who argued that these GCs followed a spherical (or at most slightly flattened) distribution
which was 
more closely concentrated to the Galactic center than their metal-poor (F-type) counterparts.
Later, Zinn (1985) and Armandroff (1989)
suggested that the metal-rich GCs belonged to the thick disk, whereas 
Minniti (1995) argued, on the basis of improved metallicities and kinematics of
bulge field stars, that most of the metal-rich GCs are in fact associated with the bulge.
Using improved distances and metallicities
for GCs within 5$^{\circ}$ of the Galactic center, Barbuy, Bica \& Ortolani (1998)
also suggested that these clusters belong to the bulge.
Burkert \& Smith (1997) have even proposed that the metal-rich GCs {\sl themselves} show evidence 
for substructuring, and suggested that at least some of these metal-rich clusters are
associated with a central Galactic bar.

Minniti (1995) considered several pieces of evidence which led him to suggest that
most MR GCs belong not to the Galactic thick disk but, rather, to the bulge.
First, and perhaps most
importantly, the metal-rich GCs are {\it highly} concentrated to the Galactic center ---
more so than would be expected if they are associated with an exponential 
thick-disk having a scale-length of $h_r \sim$ 3 kpc ($e.g.$, Ojha et al. 1996), yet
consistent with that expected for a bulge-like density distribution
($e.g.$ ${\rho}_b \propto R_G^{-m}$ where $3.65 \lae m \lae 4.2$;
Terndrup 1988, Blanco \& Terndrup 1989).
Supporting evidence for the association of the metal-rich GCs with the bulge included the
agreement between the mean metallicities for metal-rich GCs and bulge stars, and the
similarity between the morphology of the CMDs for metal-rich GCs and those for the 
bulge field population ($e.g.$, Ortolani et al. 1995b). Minniti (1995) also
considered the kinematic properties of the metal-rich GCs, although in this case the
small number of GCs in the direction of the inner Galaxy having accurate velocities 
made it impossible to draw firm conclusions. For instance, in the most recent edition of 
the Milky Way GC Catalog (see Harris 1996),\altaffilmark{2}\altaffiltext{2}{The Catalog of
Parameters for Milky Way Globular Clusters is available electronically at
http://physun.physics.mcmaster.ca/GC/mwgc.dat.} 
roughly 15\% of the GCs lack
measured radial velocities. Many of these objects are low-concentration GCs in the
direction of the inner Galaxy which suffer from heavy reddening and extreme crowding by
field stars: characteristics which have precluded radial velocity 
measurements based on integrated-light spectroscopy. 

In this paper, we report the first radial velocities for eleven GCs in
the direction of the inner Galaxy, as well as approximate metallicities for
two GCs lacking previous estimates. These data are supplemented with improved
reddenings, metallicities and distances for many of the innermost Galactic GCs
(Barbuy, Bica \& Ortolani 1998; and references therein) in order to
investigate the kinematics of the Galactic GC system and, in particular, to compare the
kinematics of the metal-rich and metal-poor GC populations to those of the
Galactic halo, disk and bulge components.

\section{Observations and Reductions}

\subsection{Target Selection}

The eleven GCs targeted in this study are listed in Table 1. Since most of these 
clusters are located in crowded fields, it was necessary to select prospective
cluster red giants based on their location in the color-magnitude diagram (CMD). 
In several cases, adequate CMDs are available in the literature
($i.e.$, NGC 6256, Alcaino 1983; NGC 6380 and Terzan 12, Ortolani, Bica \& Barbuy 
1998; Ton 2, Bica, Ortolani \& Barbuy 1996; Djorg 1, Ortolani, Bica \& Barbuy 1995a).
For the remaining GCs, no suitable data could be found in the literature, and
new multi-filter imaging was required. 

We used the Palomar 1.5m telescope and direct CCD camera with a 2048$\times$2048
detector (scale = 0\farcs37 pixel$^{-1}$) with Johnson $V$ and Kron $I$ filters to 
obtain images of Terzan 3, IC 1257, NGC 6540, IC 1276 and Pal 10 on 1-2 June 1998.
Both nights suffered from intermittent clouds, so candidate cluster members were selected
from instrumental CMDs generated using the DAOPHOT II photometry package (Stetson 1993).
For one additional cluster, NGC 6749, candidate member stars were selected from a
$V,V-I$ CMD constructed from 
archival CFHT images retrieved from the Canadian Astronomy Data Centre.
Finding charts for these six GCs are shown in Figures 1-6. All stars observed 
with HIRES are identified using the numbering scheme adopted in
Table 1. For the remaining clusters, the identification numbers
correspond to those of the photometric studies referenced in the final column of Table 1.
When available, magnitudes and colors for these stars are recorded in columns 6-8.

\subsection{HIRES Spectroscopy}

We used the High Resolution Echelle Spectrometer (HIRES; Vogt et al. 1994) on 
25-26 June 1998 to obtain high-resolution, low-S/N
spectra for 54 candidate member red giants in the eleven GCs listed in Table 1. 
The 2048$\times$ 2048 detector was binned 2$\times$2,
and the lone readout amplifier was adjusted to the high gain setting of 2.4 $e^{-1}$/ADU.
The red collimator was used along with the cross-disperser in first order. The 
latter was adjusted to an angle of $-0.2^{\circ}$ which, combined with the
adopted grating angle of 0.0$^{\circ}$, yielded a spectral coverage of 
4200 $\lae \lambda\lambda \lae$ 6600 \AA  . The C1 decker was used to limit to the entrance aperture
to 0\farcs86$\times$7\farcs0. Although conditions were photometric on both nights,
the seeing was typically $\sim$ 1\farcs2. Exposure times ranged 
from 60s to 1800s. Thorium-Argon comparison lamp spectra were taken immediately
before and after each telescope movement, or roughly every hour in those cases where 
a particular GC was observed for an extended period.

The images were trimmed, overscan-corrected and bias-subtracted in the usual manner. As a first
step in the removal of cosmic rays, the IRAF\altaffilmark{3}\altaffiltext{3}{IRAF is 
distributed by the National Optical Astronomy Observatories, which are operated by 
the Association of Universities for Research in Astronomy, Inc., under contract to 
the National Science Foundation.}
task COSMICRAY was run on the 
images. With the APALL task, six echelle orders spanning the range 4920 to 5350 \AA\ 
were then identified, traced and extracted using high-S/N spectra
of radial velocity standard stars taken at the beginning of each night. The
resulting traces were then used to extract all remaining spectra in an identical manner,
with the sole difference that the recentering option in APALL was invoked to account 
for small drifts of the object spectra perpendicular to the dispersion axis.
Th-Ar comparison spectra were extracted in the same manner, and a total 
of 22-31 emission lines per order were used to derive the best-fit coefficients of
a sixth-order Legendre polynomial used to parametrize the disperison solution. 

Object spectra were dispersion corrected with the REFSPEC and DISPCOR
tasks, using the comparison lamps which bracketed each set of observations for
a particular cluster. The rms scatter in the fitted dispersion solution
was typically $\sim$ 0.03 \AA\, with a dispersion of 4.0 km s$^{-1}$ pixel$^{-1}$
near 5175 \AA .
Cosmic rays which survived the initial cleaning of the images were then removed 
by fitting a third-order spline to the processed spectra, replacing those points
more than 4$\sigma$ above the continuum with the appropriate value of the fitted continuum.
Heliocentric radial velocities were then measured using the FXCOR task. After some experimentation,
it was found that the highest cross-correlation peaks were
found for the order spanning the region 5128-5196 \AA . 
As a result, the final radial velocity for each star represents the weighted mean of 
the derived radial velocities for this single order, using nine template spectra for
five different radial velocity standards acquired throughout the two-night run
($i.e.$, HD182572, HD186791, HD223311, HD107328 and HD90861).

The results are summarized in Table 1 which gives, for each cluster, the 
identification of all stars observed with HIRES, the exposure time and the 
measured heliocentric radial velocity, the heliocentric Julian date,
$V$ magnitude, $(B-V)$ and $(V-I)$ color for each star (if available), and 
the reference for the photometry.

\section{Sample Properties}

\subsection{New Metallicities}

Although the spectra for the majority of our program objects are of low signal-to-noise
($i.e.$, S/N $\sim$ 5 pixel$^{-1}$), a small number of stars have
S/N $\sim$ 10-15 pixel$^{-1}$. This is just adequate to measure equivalent widths for the
strongest metallic absorption lines.
The IRAF task SPLOT was therefore used to
determine equivalent widths for the relatively strong and isolated
FeI lines at 5328.051, 5216.283, 5232.952, 5171.610 and 5339.937\AA\
for 22 member stars in Pal 10, Ton 2, IC 1257, NGC 6749, NGC 6540, Terzan 3 and
IC 1276. For the last two
clusters, there are no published metallicity estimates, so we used the
empirical relation between mean equivalent width and published metallicity for Pal 10,
Ton 2, IC 1257 and NGC 6749 to
estimate [Fe/H] $\sim -0.75\pm0.25$ for both Terzan 3 and IC 1276 (see Figure 7).
The quoted uncertainty reflects the observed scatter about the adopted relation, presumably due
to neglecting the small differences in effective temperature and surface gravity
among the calibrating stars (along with the uncertainties in the published cluster
metallicities, most of which have been derived photometrically).
During this exercise, we noticed that the equivalent widths of four
stars in NGC 6540 appeared slightly lower than those expected for a cluster
having metallicity [Fe/H] = $-1.0$, the value estimated for NGC 6540 by Bica et al. (1994)
on the basis of $UBVI$ photometry and low-resolution, integrated-light spectroscopy. Given
the crowded nature of this field, it is possible that the earlier estimate
may have been biased upwards by interloping disk and bulge field stars. Based on the 
empirical relation shown in Figure 7, we estimate [Fe/H] $\simeq -1.40\pm0.25$ for NGC 6540.

\subsection{Observed and Derived Parameters for Program Clusters}

Table 2 gives a number of observed parameters for our eleven program GCs. From left to right,
this table lists the cluster name, right ascension, declination, Galactic longitude and 
latitude, integrated $V$-band magnitude, reddening, concentration index and
core radius. With the exception of the new and revised metallicities for the three 
GC discussed above, all of the data are taken from Barbuy et al. (1998) and the 
1997 edition of the Milky Way GC Catalog of Harris (1996).
Table 3 gives derived parameters for the same sample of GCs. Columns 1-7 record, from
left to right, the cluster name, heliocentric distance, Galactocentric distance, 
rectangular coordinates in the system of Harris (1996) and metallicity (which, in almost
every case, has been derived from the morphology of the observed CMD). Columns 8-10 give the number of stars 
found to be cluster members, the mean radial velocity of these stars and their intrinsic
one-dimensional velocity dispersion. These latter two quantities were calculated using
the technique described in Suntzeff et al. (1993). 

No velocity dispersion is reported for Djorg 1 and Pal 10
since only two radial velocities are available for stars in each of these clusters; 
in both cases, the measured pair of velocities are in excellent agreement,
suggesting that both stars are bonafide cluster members. Nevertheless, given the severe 
crowding in these fields, we consider the systemic radial velocities for these
clusters to be provisional, particularly that of Djorg 1 whose 
velocity with respect to a stationary observer at the location
of the Sun (see \S3.3) would be, if confirmed, the largest yet measured. Indeed, if we 
assume that Djorg 1 is currently at perigalacticon ($i.e.$ $R_p$ = 2.4 kpc) then, for a 
logarithmic Galactic potential having a circular velocity of $v_c$ = 220 km s$^{-1}$, the 
corresponding apocentric distance is $$R_a = R_p\exp{(v^2_o/2v^2_c)} \eqno{(1)}$$ 
where $v_o$ is the cluster space velocity. For ${\overline{v}}_r \le v_o \lae \sqrt{3}{\overline{v}}_r$,
the inferred apocentric distance falls in the range $10 \lae R_a \lae 140$ kpc, which places Djorg 1 in the outer halo.
This location seems incompatible with the published metallicity of [Fe/H] = $-0.4$
(Ortolani et al. 1995a) for this cluster; we note that the 
existing CMD for this cluster provides only weak constraints on
the cluster metallicity, raising the possibility that Djorg 1 may
be a misclassified MP halo cluster. Improved imaging of Djorg 1 and,
especially, spectroscopy of additional candidate members are clearly warranted.

\subsection{Provisional Mass-to-Light Ratios}

Although radial velocities are available for only a small number of stars in each GC, the
velocity precision is sufficiently high that it is possible to derive rough mass-to-light 
ratios for these GCs. Illingworth (1976) showed that the total mass for a GC can be
expressed as
$$M = 167r_c\mu\sigma^2_0 \eqno{(2)}$$ where
$r_c$ is the core radius in parsecs, $\mu$ is the dimensionless mass of the best-fit King model 
(King 1966) and $\sigma_0$ is the radial component of the central velocity dispersion. We
approximate this latter quantity with $\sigma_v$ (given in column 10 of Table 8).
This is a reasonable approximation since most of our member stars are situated in
the cores of their respective clusters. The final three columns of Table 8 record
the total luminosities, total masses and mass-to-light ratios for the nine
program clusters having three or more measured radial velocities.
These mass-to-light ratios are, of course, highly uncertain owing to the small 
number of available velocities. Nevertheless, it is reassuring to note that the weighted 
mean for the sample is $M/L_V$ = 2.2$\pm$1.1 in solar units: indistinguishable from the 
mass-to-light ratios found for nearby GCs using much larger samples of radial velocities
($e.g.$, Pryor \& Meylan 1993). 

\section{Kinematics of the Galactic Globular Cluster System}

\subsection{Parsing the Sample on the Basis of Metallicity}

An outstanding question in GC research is the origin of the chemically-distinct
GC populations associated with many giant galaxies (see, $e.g.$, Ashman \& Zepf 1992; 
Forbes, Brodie \& Grillmair 1997; C\^ot\'e, Marzke \& West 1998). Central to this
question are the issues of substructuring in the Galactic GC system and the proper 
categorization of the various GC subsystems with the Galactic
halo, bulge/bar or thick disk. We regard the bimodal metallicity distribution
of Galactic GCs the strongest single piece of evidence for the presence of
distinct GC populations in the Milky Way. Thus, we proceed by first dividing 
the Galactic GC system into metal-rich and metal-poor components, and then analyzing
their kinematics separately.

Figure 8 shows the metallicity distribution of 133 Galactic GCs having measured
chemical abundances. The metallicities --- which are essentially on the Zinn \& West (1984) 
scale ---  are taken from the most recent version of the catalog 
of Harris (1996),
supplemented with recent data from Barbuy et al. (1998) and with the metallicity
estimates for Terzan 3, IC1276 and NGC 6540 presented in \S 3.1.
The evidence for two distinct GC populations, first suggested by Morgan (1959), is
immediately apparent as two peaks in the distribution.
The double-Gaussian which best fits the observed data is 
shown by the solid line; the metal-poor (MP) and metal-rich (MR) components are
indicated by the dotted lines. These Gaussians have parameters
$N_{\rm mp} = 14.9$ per 0.125 dex, [Fe/H]$_{\rm mp} = -1.59$ dex and ${\sigma}_{\rm mp} = 0.30$ dex
for the MP component, and $N_{\rm mr} = 7.1$ per 0.125 dex, [Fe/H]$_{\rm mr} = -0.55$ dex and 
${\sigma}_{\rm mr} = 0.27$ dex for the MR component.

Most previous discussions concerning the proper categorization of these two 
GC populations have involved dividing the sample into two components at some 
intermediate metallicity. However, it is clear from Figure 8 that approach is
likely result in the inclusion of some GCs in the MR component which are more
properly associated with the MP GC population, and vice versa.
In an attempt to avoid such complications, we choose to divide the sample into
{\sl three} components: (1) a sample of 31 MR GCs (including Djorg 1) having [Fe/H] $< -0.75$; (2)
a sample of 82 MP GCs having [Fe/H] $\le -1.29$ and; (3) a sample of 
20 intermediate metallicity (IM) GCs having $-1.29 <$ [Fe/H] $\le -0.75$.
Thus, although a mixture of MP and MR GCs are expected in the IM sample,
the MR and MP samples should be relatively uncontaminated by interlopers.

\subsection{Parametric Solutions for Velocity Dispersion and Rotational Velocity}

We now investigate the kinematics of these three GC samples using familiar parametric 
techniques first described in Kinman (1959), Frenk \& White (1980), Zinn (1985)
and Hesser, Shawl \& Meyer (1986).
First, we consider the case of constant rotation velocity, $v_r$.
A population of GCs having constant $v_r$ should fall along a relation of the form
$$v_s = v_{r}\cos{\psi} + v_p\eqno{(3)}$$
where $v_s$ is the velocity with respect to a stationary
observer at the location of the Sun, $\psi$ is the angle between
the rotational-velocity vector of the GC and the direction to the Sun,
and $v_p$ is the peculiar velocity of the GC with respect to the derived
solution. The angle $\psi$ is given by
$$\cos{\psi} = {R_{\odot}\cos{A} \over \sqrt{R_s^2\cos{A}^2 +(R_{\odot}-R_s\cos{b}\cos{l})^2}}\eqno{(4)}$$
where $R_{\odot}$ is the Galactocentric distance of the Sun (taken to be 8 kpc), 
$R_s$ is the heliocentric distance of the GC, and $l$ and $b$ are its Galactic
longitude and latitude.\altaffilmark{4}\altaffiltext{4}{Note that equation (2) of
Burkert \& Smith (1997) contains a typographical error. In their notation, $R$ 
should denote the {\sl heliocentric} distance of the cluster.} 
The angle $A$ is the angle between the
apex of the Local Standard of Rest (LSR) and the position of the GC on the sky,
given by $\cos{A} = \sin{l}\cos{b}$. Thus, $v_s = v_{\rm lsr} + 220\cos{A}$ where $v_{\rm lsr}$
is the velocity with respect to the LSR.

A second, frequently-used rotation law is the case of solid-body rotation. In this
instance, $v_s$ is given by
$$v_s = {\omega}{R_{\odot}}\cos{A} + v_p\eqno{(5)}$$
where $\omega$ is the angular velocity. The observed flat rotation curve of the thick disk
(Ojha et al. 1996) suggests that equation 3 should provide an adequate description for the
observed radial velocities of thick disk objects, although it is important to bear in mind
that the available data are limited to the solar neighborhood ($i.e.$, there exist virtually no
observational constraints on the kinematics of thick-disk stars within a few
kpc of the Galactic center).
On the other hand, equation 4 should give a superior 
representation for objects belonging to the Galactic bulge --- an oblate isotropic rotator 
which is adequately described by solid-body rotation (Kent 1992; Ibata \& Gilmore 1995; 
Tiede \& Terndrup 1997). 
%It is also worth bearing in mind that, although equation 3 should
%closely reflect the kinematics of young disk stars, its applicability to the
%{\sl thick} disk is essentially unknown ($i.e.$, there exist virtually no 
%observational constraints on the kinematics of thick-disk stars within a few
%kpc of the Galactic center).

We have derived the best-fit values of $v_r$, $\omega$, and their associated 
line-of-sight velocity dispersions, $\sigma_v$ and $\sigma_{{\omega}R_{\odot}}$, using
various sub-samples of the Galactic GC population.
The results of this exercise is summarized in Table 4, whose first four columns 
record the metallicity of the chosen GC sample ($i.e.$, MR, MP or IM), the 
range of Galactocentric radii spanned by these GCs, their mean Galactocentric radius, 
and the total number of objects in the sample. Columns 4-8 list
the calculated values of $v_r$, $\sigma_v$, $\omega$, $\sigma_{\omega}$ and 
the ratio of ordered-to-random motions, $v_r / \sigma_v$. The
number given in the final column identifies the sample used in calculating 
the best-fit kinematic parameters. In addition to dividing the sample on the
basis of metallicity, we have parsed the sample on the basis of Galactocentric
radius by considering three different cases: (1) GCs interior to $R_G = 4$ kpc; (2) GCs 
beyond $R_G = 4$ kpc and; (3) all GCs, irrespective of distance from the Galactic
center. Note that this choice
of dividing radius, though arbitrary, corresponds roughly to the co-rotation radius
of the Galactic bar observed in 3D N-body simulations ($e.g.$, 3-5 kpc; Fux 1997), the effective radius of the
Galactic bulge (2.7 kpc; Gilmore, King \& van der Kruit 1990) and the median $K$-band
effective radius measured for the bulges of Sb and Sbc galaxies
($i.e.$, $\simeq$ 4 kpc for $H_0$ = 75 km s$^{-1}$ Mpc$^{-1}$; de Jong 1996).
Roughly two-thirds (20/31) of the MR GCs (which span the range $0.6 \le R_G \le 7.6$ kpc with mean
${\overline{R_G}} = 3.2\pm2.0$ kpc) are located within 4 kpc of the Galactic center.

The results of this exercise are shown in Figures 9 and 10.
The distribution of $v_s$ with $\cos{\psi}$ for Galactic GCs
is shown in Figure 9, whereas Figure 10 shows the distribution of $v_s$ with $\cos{A}$. 
While most of the data used to construct these figures have been taken
from of Harris (1996), we have added new 
velocities for the eleven GCs studies here and have
made use of the improved radial velocities of seven outer halo GCs
given in Djorgovski et al. (1999).
In both figures, the upper panel shows the distribution for the MR GCs (circles);
MP and IM GCs are plotted in the lower panel (squares and triangles, respectively).
As described above, all three samples have been sub-divided on the basis of 
Galactocentric radius;
open symbols refer to GCs located more than 4 kpc from the Galactic center,
whereas GCs interior to this radius are plotted as filled symbols. 

Also shown in Figures 9 and 10 are the best-fit relations of the form of equation
3 and 5, respectively. Each line is marked with the number given in the last 
column of Table 4, identifying the sample of GCs used in deriving the 
rotation solution (except for the lower panel of Figure 9 where the labels
have been omitted for clarity). A number of conclusions may be drawn from these 
figures. First, the MR GCs show significant rotation at all 
radii.\altaffilmark{5}\altaffiltext{5}{We omit Djorg 1 from the MR sample since
its radial velocity and metallicity are uncertain; including it increases
the derived rotation velocity and line-of-sight velocity dispersion for the sample 
of inner MR GCs by $\sim$ 15\% and 20\%, respectively (see Table 4).}
Second, only among the outermost MR GCs does the quantity $v_r/ \sigma_v$ substantially 
exceed unity, meaning that this component of the Galactic GC system is supported 
primarily by ordered, and not random, motions (Zinn 1985; Armandroff 1989).
Third, the MP GCs as a whole show little or no evidence for rotation:
a familiar result which is well established from earlier studies ($e.g.$, Kinman 1959;
Frenk \& White 1980; Zinn 1985; Hesser et al. 1986) although we note that there is some indication 
that the inner sample of MP GCs {\sl may} be rotating slowly. Not surprisingly, the results for 
the IM sample are intermediate to those found for the MR and MP GCs: $i.e.$, rotation 
is generally significant at all radii, but of lower
amplitude that for the sample of MR GCs. This is not unexpected since the sample IM
GCs almost certainly represents an inhomogeneous mixture of the MR and MP samples.

\subsection{Comparison to the Bulge Field Star Population}

Is the observed rotation of MR GCs in the inner Galaxy more characteristic
of the bulge or disk populations? The dashed lines in the upper and lower
panels of Figure 11 show the bulge rotation law found by Tiede \& Terndrup 
(1997) over the range $-10^{\circ} \le l \le 10^{\circ}$.
The dotted lines which span the range $5^{\circ} \le |l| \le 25^{\circ}$ show the rotation
law for the outer bulge measured by Ibata \& Gilmore (1995) for $|b| \ge 12^{\circ}$,
offset by $\pm$ 55 km s$^{-1}$ in order to match the 
relation of Tiede \& Terndrup (1997) over the range $5^{\circ} \lae |l| \lae 10^{\circ}$.
MR GCs are plotted in the upper panel; MP and IM GCs are given
in the lower panel. As in the previous two figures, open symbols refer to
GCs having $R_G > 4$ kpc, while the filled symbols indicate objects with 
$R_G < 4$ kpc.  The errorbar in the upper panel shows the typical velocity 
dispersion of 110 km s$^{-1}$ for bulge objects, as summarized in Table 1 of Kent (1992).
This dispersion agrees closely with that found for the MR GCs in \S 4.2, particularly
for those clusters with $R_G \le 4$ kpc. In addition, 
the agreement between the rotational laws for bulge stars and the MR
GCs appears quite reasonable. Interestingly, the innermost MP and IM GCs also show
reasonable agreement with rotation law deduced from observations of bulge field stars. 

\subsection{Comparison to the HI Gas in the Inner Galaxy: Is there Evidence for Bar-like Kinematics?}

Unfortunately, a direct comparison to the kinematics of the thick-disk field star 
population in the central few kiloparsecs of the Galaxy is not possible, as there exist 
no observational constraints on the kinematics of the thick-disk in this region. 
In fact, there are surprisingly few constraints on the kinematics of the {\sl old thin disk} 
in the inner Galaxy, although it is thought that its velocity dispersion increases 
substantially within the central few kiloparsecs ($e.g.$, Lewis \& Freeman 1989).

However, there do exist excellent observations of molecular and 
atomic gas in the inner Galaxy whose kinematics can be compared directly to that 
of the GCs. Figure 12 shows the well known longitude-velocity diagram for
HI gas in the inner Galaxy based on 21 cm observations ($e.g.$ Burton \& Liszt 1993).
Detailed discussions of the potential pitfalls involved in interpreting the gas 
kinematics may be found in Burton \& Liszt (1993)
and List (1992). Here, we simply summarize the two principal implications of these observations
for Galactic structure models: (1) gas motions in the inner Galaxy
deviate strongly from circularity; and (2) the origin of the non-circular
motions may arise from the presence of a bar or a triaxial bulge potential 
in the inner Galaxy. The existence of a Galactic bar is now supported by
much additional observational evidence, particularly COBE/DIRBE surface
brightness maps of the inner Galaxy (Dwek et al. 1995; Binney, Gerhard \& Spergel 1997),
the asymmetric distribution of bulge red clump stars (Stanek et al. 1994; 1997) and the
non-circular kinematics of $^{12}$CO, $^{13}$CO and CS gas (Bally et al. 1987; 
Dame et al. 1987; Binney et al. 1991).

The first claims of a bar-like (or highly flattened disk-like) feature in the spatial 
distribution of GCs were made by Baade (1958), Kinman (1959) and Morgan (1959). 
Woltjer (1975) and Harris (1976), however, cautioned that such a distribution might be due to the
smearing effects of distance errors for low-latitude clusters in the direction of the 
Galactic center. Recently, the possibility that at least some of the MR GCs might be associated with a
Galactic bar has been discussed by Burkert \& Smith (1997) and Barbuy et al. (1998)
who, using different samples, came to different conclusions. Using data from Harris
(1996), Burkert \& Smith (1997) argued that the highest-mass MR GCs fall
along an elongated bar whose extent and orientation is consistent with that of the
Galactic stellar bar. Barbuy et al. (1998), on the other hand, compared the spatial
distribution of GCs found within 5$^{\circ}$ of the Galactic center to the outer contours of
the Galactic bar proposed by Blitz \& Spergel (1991), and concluded that the innermost GCs show
no clear evidence for a bar. They note, however, that the solid angle spanned
by their sample of clusters is sufficiently small that a bar-like distribution might be 
undetectable. 

The velocity distribution of HI gas in the inner Galaxy, summed over the range 
$|b| \le 1.5^{\circ}$, is shown by the contours in
Figure 12. The innermost MR GCs (circles), MP GCs (squares)
and IM GCs (triangles) are overlaid for comparison. Although the combined
GC sample spans the range $|b| \lae 23^{\circ}$
(with mean $\overline{b} = 0.7^{\circ}\pm9.2^{\circ}$),
there is a striking
agreement between the kinematics of the GCs and the atomic gas. Perhaps surprisingly, this
is true of the entire sample of GCs, irrespective of metallicity. In the region of 
Galactic longitude spanned by the HI observations, there is only a single GC having
$R_G < 4$ kpc which falls outside of the envelope defined by the 
HI gas. This object (located at $l \simeq 8$ and $v_{\rm lsr} \simeq$ 200 km s$^{-1}$) is
NGC 6144: a poorly-studied GC whose Galactocentric distance of 3.4 kpc is based
on a single photographic CMD (Alcaino 1980). An improved estimate of the distance of 
this cluster is desirable. Finally, we note that the pattern speed of the Galactic
stellar bar, according to the dynamical models of Zhao (1996), is $\Omega_{\rm p}$ $\sim$ 
60 km s$^{-1}$ kpc$^{-1}$. To within the rather large uncertainties, this value is
consistent with the angular velocity of $\omega$ = 43.6$\pm$22.3 km s$^{-1}$ kpc$^{-1}$
found for the MR GCs within 4 kpc of the Galactic center.

\subsection{Whence the Metal-Rich Globular Clusters?}

The similarity between the longitude-velocity diagram of GCs and HI gas shown in Figure 12 
is intriguing, but does it necessarily imply that the GCs are associated with the Galactic bar? 
Although it is notoriously difficult to distinguish motion along a dynamically cold bar from
motion within a rigidly-rotating, elongated bulge having a significant velocity 
dispersion, one possible observational test of these different scenarios 
would be the existence of a distance-velocity relation among GCs: $i.e.$, if the
GCs are indeed associated with the Galactic bar, then the position of each GC in
the X-Y plane of the Galaxy should correspond uniquely to an observed radial velocity,
something which is {\sl not} true should they belong to a triaxial bulge. Figure 13
shows the Galactocentric XY coordinates of the all GCs within 4 kpc of the Galactic center.
The large circle
indicates the approximate co-rotation radius of the Galactic bar (Fux 1997), while
the dotted lines show the limits placed on the orientation of the bar by observations of
red clump giants (Stanek et al. 1997).
In this figure, open
symbols indicate clusters having $v_{\rm lsr} \ge 0$ km s$^{-1}$ while filled symbols
refer to GCs with $v_{\rm lsr} < 0$ km s$^{-1}$. The bar-like feature 
noted by earlier workers is readily apparent, although it is clear that the
XY positions of the GCs do not map uniquely into positive or negative velocities, as might
be expected in bar models (see, $e.g.$, Figure 12 of Roberts et al. 1979). It is important
to bear in mind, however, that the three-dimensional Galactocentric positions for many of
these GCs remain sufficiently uncertain that definite conclusions regarding their possible 
association
with the Galactic bar are premature. Improved distances for these heavily-obscured GCs,
preferably based on deep infrared CMDs ($e.g.$, Kuchinski et al. 1995; Minniti et al. 1995), 
are needed to settle the issue.

The situation can be summarized as follows. Based on the evidence for
rotation among the inner MR GCs, the close agreement between the overall 
rotation of the MR GC system and that of the bulge field stars, and the
agreement between the kinematics of the innermost GCs and that of the HI gas, 
we suggest that majority of the MR GCs within 4 kpc of the Galactic center are 
more properly categorized as members of the Galactic bulge and/or bar (and not
the thick disk), as suggested by Harris (1976) and re-emphasized by
Minniti (1995). We caution, however, that this
simple interpretation ($i.e.$, the association of the innermost MR GCs with the
bulge/bar) need not have been true at all times.
For instance, some dynamical models for the formation of galactic bulges suggest
that disk formation predates that of the bulge. Such models include bulge formation 
during the dissolution of bars which themselves form via disk instabilities (Hohl 1971) and 
the growth of bulges through the accretion of satellites onto disk galaxies (see, $e.g.$, 
Pfenniger 1993). Although the extremely old ages measured for MR clusters and bulge
field stars ($e.g.$ Ortolani et al. 1995b; Frogel 1998) may present difficulties for models in which
the present population of MR GCs were initially associated with the inner regions of
a disk system which ultimately became bar-unstable, additional observations (particularly
improved ages for the innermost MR GCs and the thick-disk/bulge field star populations) are
needed before such scenarios can be confirmed or ruled out.

\section{Summary}

We have presented the first radial velocities for eleven, heavily-reddened GCs in 
the direction of the inner Galaxy. For two of these clusters (Pal 10 and Djorg 1), 
the velocities should be considered provisional due to the small number of stars observed.
From the HIRES spectra used to measure these velocities, we estimate [Fe/H] $\simeq -0.75$
for Terzan 3 and IC 1276: two clusters lacking previous metallicity estimates.
We have estimated provisional mass-to-light ratios for nine of the GCs; the mean 
of $\overline{M/L_V} = 2.2\pm1.1M_{\odot}/L_{V,{\odot}}$ is similar to that found for 
nearby, well-studied clusters ($e.g.$, Pryor \& Meylan 1993).

Using an updated catalog of GC positions, distances, metallicities and radial velocities
we have investigated the kinematic properties of the MR and MP components of the Galactic
GC system. The outermost MP GCs, and the MP GC population as a whole, shows little or no 
evidence for significant rotation, as expected for a system of halo objects.
On the other hand, there may be some evidence that the innermost MP GCs are rotating slowly. 
We find that the MR GCs exhibit significant rotation at all radii, although 
rotation is dynamically important only among the outermost MR GCs.
The rotation curve and line-of-sight velocity dispersion for the MR GCs within
$\simeq$ 4 kpc of the Galactic center shows
good agreement with that observed for bulge stars. A comparison
of the longitude-velocity diagram for all GCs with $R_G \lae$ 4 kpc
shows a remarkable similarity to that of the HI gas in the inner Galaxy, including the
same non-circular motions which are often interpreted as evidence for a Galactic bar, or
alternatively, an non-axisymmetric bulge. When improved three-dimensional Galactocentric 
positions become available for many GCs in the inner Galaxy, it may be possible to 
distinguish between these competing scenarios.

\acknowledgments

The author wishes to thank Wal Sargent for allocating two nights of unassigned Keck 
time to this project, George Djorgovski for his assistance during the observing run, 
Harvey Liszt for providing the HI data shown in Figure 12, and Steve
Vogt for building such a magnificent instrument. Thanks also to John Blakeslee,
Bill Harris, Sterl Phinney and Bob Zinn for
helpful comments on earlier versions of this paper.
The author gratefully acknowledges support provided by the Sherman M. Fairchild Foundation.
This study has made use of the Canadian Astronomy Data Centre, which is operated by the 
Herzberg Institute of Astrophysics, National Research Council of Canada.

\begin{deluxetable}{lrrrrrrrc}
\tablecolumns{9}
\tablewidth{0pc}
\tablecaption{Radial Velocities for Globular Cluster Red Giant Candidates\label{tbl-2}}
\tablehead{
\colhead{Cluster} &                      
\colhead{Star} &
\colhead{$T$} &
\colhead{$v_r$} &
\colhead{HJD} &
\colhead{$V$} &     
\colhead{$(V-I)$} &                      
\colhead{$(B-V)$} &
\colhead{Reference} \\
\colhead{} &
\colhead{} &
\colhead{(sec)} &
\colhead{(km s$^{-1}$)} &
\colhead{(2440000+)} &
\colhead{(mag)} &
\colhead{(mag)} &
\colhead{(mag)} &
\colhead{} 
}
\startdata
Terzan 3   &  1 &180& -133.47$\pm$0.20 & 10990.7699 & & & & \nl
           &  3 &240& -137.43$\pm$0.19 & 10990.7813 & & & & \nl
           &  4 &240& -136.05$\pm$0.18 & 10990.7861 & & & & \nl
           &  5 &240& -133.70$\pm$0.23 & 10990.7939 & & & & \nl
           &  6 &240& -138.23$\pm$0.29 & 10990.7899 & & & & \nl
           &  7 &420& -137.94$\pm$0.23 & 10990.7992 & & & & \nl
           &  8 &420&   20.30$\pm$0.19 & 10990.8052 & & & & \nl
           &  9 &420& -138.60$\pm$0.24 & 10990.8113 & & & & \nl
           & 10 &420& -136.19$\pm$0.25 & 10990.8173 & & & & \nl
NGC 6256   & 1-5 & 90& -100.53$\pm$0.47 & 10990.8722 &15.29& &2.43& 1\nl
           & 1-8 &150& -104.25$\pm$0.39 & 10990.8748 &15.71& &2.54& 1\nl
           & 1-11&150&  -92.17$\pm$0.38 & 10990.8783 &15.62& &2.63& 1\nl
           & 1-14&150& -102.49$\pm$0.38 & 10990.8695 &15.66& &2.49& 1\nl
           & 1-13&210& -101.65$\pm$0.36 & 10990.8819 &15.94& &2.43& 1\nl
           & 1-39&300&  -96.72$\pm$0.37 & 10990.8861 &16.20& &2.15& 1\nl
           & 1-20&420& -101.33$\pm$0.34 & 10990.8914 &16.23& &2.36& 1\nl
&13\tablenotemark{a}&420&-98.83$\pm$0.31 & 10990.8984 &     & &    & \nl
           & 1-19&420&  -82.36$\pm$1.29 & 10990.9043 &15.72& &1.89& 1\nl
IC 1257    &    1 &1000& -141.79$\pm$0.40 & 10991.7681 & & & & \nl
           &    2 & 750& -137.64$\pm$0.95 & 10991.7823 & & & & \nl
           &    3 & 900& -131.41$\pm$1.03 & 10991.7935 & & & & \nl
           &    4 &1200& -140.93$\pm$1.30 & 10991.8266 & & & & \nl
           &    5 &1500& -140.91$\pm$1.11 & 10991.8092 & & & & \nl
NGC 6380   &  977 &1800&   -0.01$\pm$0.87 & 10991.8532 & 18.30 & 2.93 & & 2\nl
           &  894 &1500&   -0.42$\pm$0.69 & 10990.9304 & 18.33 & 2.72 & & 2\nl
           & 1066 &1500&   -7.24$\pm$0.56 & 10990.9507 & 18.28 & 2.78 & & 2\nl
Ton 2      & 3258 &900& -181.75$\pm$0.70 & 10990.9952 & 17.47 & 2.77 & & 3\nl
           & 3633 &900& -187.81$\pm$0.34 & 10990.9709 & 17.03 & 3.17 & & 3\nl
           & 3644 &900& -181.10$\pm$0.37 & 10990.9833 & 17.13 & 3.00 & & 3\nl
Djorg 1    & 330 &1800& -364.30$\pm$0.69  & 10991.9352  & 18.26 & 3.40 & &4\nl
           & 372 &1800& -357.63$\pm$1.09  & 10991.8892  & 18.58 & 3.62 & &4\nl
NGC 6540   & 12  &60& -21.08$\pm$0.48   & 10991.0073  & & & & \nl
           & 16  &90& -20.19$\pm$0.44   & 10991.0096  & & & & \nl
           & 27  &120& -12.73$\pm$0.41   & 10991.0221  & & & & \nl
           & 31  &90& -21.43$\pm$0.70   & 10991.0144  & & & & \nl
           & 57  &120& -18.14$\pm$0.48   & 10991.0194  & & & & \nl
           & 62  &90& -16.96$\pm$0.33   & 10991.0120  & & & & \nl
IC 1276    &  2  &120&  157.02$\pm$0.42  & 10991.0283  & & & & \nl
           &  3  &210&  153.56$\pm$0.27  & 10991.0314  & & & & \nl
           &  4  &270&  152.30$\pm$0.32  & 10991.0353  & & & & \nl
           &  5  &360&  155.15$\pm$0.23  & 10991.0400  & & & & \nl
           &  6  &420&  -12.16$\pm$0.24  & 10991.0458  & & & & \nl
           & 10  &420&  159.86$\pm$0.25  & 10991.0518  & & & & \nl
Terzan 12  & 294 &1800&   91.70$\pm$0.68  & 10992.0617  & 18.95 & 3.94 & & 2\nl
           & 410 &1200&   95.19$\pm$0.66  & 10992.0235  & 18.33 & 4.39 & & 2\nl
           & 428 &1500&   96.75$\pm$1.01  & 10992.0407  & 18.75 & 4.25 & & 2\nl
NGC 6749   &  1  &300& -60.55$\pm$0.47   & 10992.0787  & & & & \nl
           &  2  &300& -60.00$\pm$1.05   & 10992.0836  & & & & \nl
           &  3  &420& -56.60$\pm$1.39   & 10992.0891  & & & & \nl
           &  4  &750& -71.36$\pm$0.76   & 10992.0973  & & & & \nl
           &  5  &750& -54.85$\pm$0.94   & 10992.1072  & & & & \nl
           &  6  &750&  52.01$\pm$0.57   & 10992.1170  & & & & \nl
Pal 10     &  1  &1500& -31.21$\pm$0.35   & 10991.1105  & & & & \nl
           &  5  &1500& -32.09$\pm$0.31   & 10991.0922  & & & & \nl
\enddata

\tablenotetext{a}{Star located 8\farcs6 south and 6\farcs2 west of star 1-20.
REFERENCES:
(1) Alcaino 1983;
(2) Ortolani, Bica \& Barbuy 1998;
(3) Bica, Ortolani \& Barbuy 1996;
(4) Ortolani, Bica \& Barbuy 1995a.
}

\end{deluxetable}

\clearpage

\begin{deluxetable}{lrrrrrccc}
\tablecolumns{9}
\tablewidth{0pc}
\tablecaption{Observed Parameters for Program Clusters\label{tbl-2}}
\tablehead{
\colhead{Cluster} &
\colhead{$\alpha$(2000)} &
\colhead{$\delta$(2000)} &
\colhead{$l$} &
\colhead{$b$} &
\colhead{$V_T$} &
\colhead{$E(B-V)$} &
\colhead{$c$} &
\colhead{$r_c$}  \nl
\colhead{} &
\colhead{(h m s)} &
\colhead{($\circ$ $\prime$ $\prime\prime$)} &
\colhead{(deg)} &
\colhead{(deg)} &
\colhead{(mag)} &
\colhead{(mag)} &
\colhead{} &
\colhead{(arcmin)}
}
\startdata
Terzan 3   &16:28:40.1&-35:21:13&345.1 &  9.2 &12.00&0.32 &0.70&1.18\nl
NGC 6256   &16:59:32.6&-37:07:17&347.8 &  3.3 &11.29&0.84 &2.50&0.02\nl
IC 1257    &17:27:08.5&-07:05:35& 16.5 & 15.1 &13.10&0.73 &2.24&0.05\nl
NGC 6380   &17:34:28.0&-39:04:09&350.2 & -3.4 &11.31&1.07 &1.55&0.34\nl
Ton 2      &17:36:10.5&-38:33:12&350.8 & -3.4 &12.24&1.26 &1.30&0.54\nl
Djorg 1    &17:47:28.3&-33:03:56&356.7 & -2.5 &13.60&1.70 &1.50&0.50\nl
NGC 6540   &18:06:08.6&-27:45:55&  3.3 & -3.3 & 9.30&0.60 &2.50&0.03\nl
IC 1276    &18:10:44.2&-07:12:27& 21.8 &  5.7 &10.34&0.92 &1.29&1.08\nl
Terzan 12  &18:12:15.8&-22:44:31&  8.4 & -2.1 &15.63&2.06 &0.57&0.83\nl
NGC 6749   &19:05:15.3&+01:54:03& 36.2 & -2.2 &12.44&1.50 &0.83&0.77\nl
Pal 10     &19:18:02.1&+18:34:18& 52.4 &  2.7 &13.22&1.66 &0.58&0.81\nl
\enddata

\end{deluxetable}

\begin{deluxetable}{lrrrrrccrrrrc}
\tablecolumns{13}
\tablewidth{0pc}
\scriptsize
\tablecaption{Derived Parameters for Program Clusters\label{tbl-2}}
\tablehead{
\colhead{Cluster} &
\colhead{$R_{\odot}$} &
\colhead{$R_G$} &
\colhead{$X$} &
\colhead{$Y$} &
\colhead{$Z$} &
\colhead{[Fe/H]} &
\colhead{$N_*$} &
\colhead{${\overline{v}}_r$} &
\colhead{$\sigma_v$} &
\colhead{$L_V$} &
\colhead{$M$} &
\colhead{$M/L_V$} \nl
\colhead{} &
\colhead{(kpc)} &
\colhead{(kpc)} &
\colhead{(kpc)} &
\colhead{(kpc)} &
\colhead{(kpc)} &
\colhead{(dex)} &
\colhead{} &
\colhead{(km s$^{-1}$)} &
\colhead{(km s$^{-1}$)} &
\colhead{(10$^3$$L_{{\odot},V}$)} &
\colhead{(10$^3$$M_{\odot}$)} &
\colhead{($M_{\odot}/L_{{\odot},V}$)} 
}
\startdata
Terzan 3   &26.4 & 18.9 &-17.2 & -6.7 &  4.2 & -0.75 &8& -136.3$\pm$0.7 & 1.9$\pm$0.5 &23.6 & 28.8 &   1.2$\pm$0.6 \nl
NGC 6256   & 9.3 &  2.3 & -1.1 & -2.0 &  0.5 & -1.01 &9&  -99.5$\pm$2.4 & 4.1$\pm$1.0 &24.9 & 30.3 &   1.2$\pm$0.6 \nl
IC 1257    &24.5 & 17.3 &-14.6 &  6.7 &  6.4 & -1.70 &5& -140.2$\pm$2.1 & 3.4$\pm$1.1 &23.6 & 83.7 &   3.5$\pm$2.2 \nl
NGC 6380   & 9.8 &  2.4 & -1.6 & -1.7 & -0.6 & -0.50 &3&   -3.6$\pm$2.5 & 4.2$\pm$1.7 &52.0 & 83.0 &   1.6$\pm$1.3 \nl
Ton 2      & 6.4 &  2.0 &  1.7 & -1.0 & -0.4 & -0.60 &3& -184.4$\pm$2.2 & 4.0$\pm$1.6 &16.3 & 46.9 &   2.9$\pm$2.4 \nl
Djorg 1    & 5.6 &  2.4 &  2.4 & -0.3 & -0.2 & -0.40 &2& -362.4$\pm$3.6 &             &12.5 &      &               \nl
NGC 6540   & 3.0 &  5.0 &  5.0 &  0.2 & -0.2 & -1.40 &6&  -17.7$\pm$1.4 & 3.1$\pm$0.9 & 8.2 &  8.7 &   1.1$\pm$0.6 \nl
IC 1276    & 9.3 &  3.6 & -0.6 &  3.5 &  0.9 & -0.75 &5&  155.7$\pm$1.3 & 3.0$\pm$1.0 &74.5 & 75.5 &   1.0$\pm$0.6 \nl
Terzan 12  & 3.4 &  4.6 &  4.6 &  0.5 & -0.1 & -0.50 &3&   94.1$\pm$1.5 & 2.3$\pm$0.9 & 2.1 &  2.9 &   1.4$\pm$1.1 \nl
NGC 6749   & 7.7 &  4.9 &  1.8 &  4.6 & -0.3 & -1.60 &5&  -61.7$\pm$2.9 & 5.7$\pm$1.8 &38.7 & 63.3 &   1.6$\pm$1.0 \nl
Pal 10     & 5.8 &  6.4 &  4.5 &  4.6 &  0.3 & -0.10 &2&  -31.7$\pm$0.4 &             &16.9 &       &               \nl
\enddata

\end{deluxetable}
\clearpage

\begin{deluxetable}{lcrrrrrrrc}
\tablecolumns{10}
\tablewidth{0pc}
\tablecaption{Rotation Solutions for Galactic GC Subsamples\label{tbl-2}}
\tablehead{
\colhead{Sample} &
\colhead{$R_G$} &
\colhead{${\overline{R_G}}$} &
\colhead{$N$} &
\colhead{$v_r$} &
\colhead{$\sigma_v$} &
\colhead{$\omega$} &
\colhead{$\sigma_{\omega{R_{\odot}}}$} &
\colhead{$v_r/ \sigma_v$} &
\colhead{ID} \nl
\colhead{} &
\colhead{(kpc)} &
\colhead{(kpc)} &
\colhead{} &
\colhead{(km s$^{-1}$)} &
\colhead{(km s$^{-1}$)} &
\colhead{(km s$^{-1}$ kpc$^{-1}$)} &
\colhead{(km s$^{-1}$)} &
\colhead{} &
\colhead{}  \nl
}
\startdata
MR\tablenotemark{a}&$>$0   & 3.2 & 30 &105$\pm$28 &100 & 30.3$\pm$\tg6.8  & 94 &1.0& 1 \nl
                   &0-4    & 1.9 & 19 & 63$\pm$41 &110 & 43.6$\pm$22.3    &107 &0.6& 2 \nl
                   &$\ge$4 & 5.4 & 11 &157$\pm$28 & 67 & 28.5$\pm$\tg5.3  & 70 &2.3& 3 \nl
&&&&&&&&&\nl
MP                 &$>$0   &16.2 & 82 & 40$\pm$26 &124 &  4.7$\pm$\tg3.7  &125 &0.3& 4 \nl
                   &0-4    & 2.3 & 23 & 68$\pm$43 &128 & 19.1$\pm$20.8    &133 &0.5& 5 \nl
                   &$\ge$4 &21.6 & 59 & 22$\pm$33 &123 &  4.2$\pm$\tg3.7  &122 &0.2& 6 \nl
&&&&&&&&&\nl
IM                 &$>$0   & 5.9 & 20 & 79$\pm$38 &100 & 11.6$\pm$\tg8.9  &106 &0.8& 7 \nl
                   &0-4    & 2.1 & 11 & 87$\pm$41 & 88 & 41.7$\pm$16.5    & 83 &1.0& 8 \nl
                   &$\ge$4 &10.6 &  9 & 64$\pm$76 &118 &  5.0$\pm$11.3    &121 &0.5& 9 \nl
&&&&&&&&&\nl
\cline{1-9}\nl
MR\tablenotemark{b}&$>$0   & 3.2 & 31 &110$\pm$32 &117 & 31.1$\pm$\tg8.1  &113 &0.9&  \nl
                   &0-4    & 2.0 & 20 & 72$\pm$50 &135 & 50.5$\pm$27.2    &130 &0.5&  \nl
                   &$\ge$4 & 5.4 & 11 &157$\pm$28 & 67 & 28.5$\pm$\tg5.3  & 70 &2.3&  \nl
&&&&&&&&&\nl
\enddata
\tablenotetext{a}{Excluding Djorg1}
\tablenotetext{b}{Including Djorg1}
\end{deluxetable}
\clearpage

\clearpage

\figcaption[gcskin.01.ps]{V-band finding chart for Terzan 3. This image measures
6\farcm3$\times$6\farcm3. North is up and east is to the left. The numbering scheme
is the same as that of Table 1.
\label{fig1}}

%\plotone{gcskin.02.ps}

\figcaption[gcskin.02.ps]{V-band finding chart for IC 1257. This image measures
1\farcm5$\times$1\farcm5. North is up and east is to the left. The numbering scheme
is the same as that of Table 1.
\label{fig2}}

%\plotone{gcskin.03.ps}

\figcaption[gcskin.03.ps]{V-band finding chart for NGC 6540. This image measures
1\farcm5$\times$1\farcm5. North is up and east is to the left. The numbering scheme
is the same as that of Table 1.
\label{fig3}}

%\plotone{gcskin.04.ps}

\figcaption[gcskin.04.ps]{V-band finding chart for IC 1276. This image measures
3\farcm0$\times$3\farcm0. North is up and east is to the left. The numbering scheme
is the same as that of Table 1.
\label{fig4}}

%\plotone{gcskin.05.ps}

\figcaption[gcskin.05.ps]{V-band finding chart for NGC 6749. This image,
taken from the CFHT acrhive, measures
1\farcm5$\times$1\farcm5. North is up and east is to the left. The numbering scheme
is the same as that of Table 1.
\label{fig5}}

%\plotone{gcskin.06.ps}

\figcaption[gcskin.06.ps]{V-band finding chart for Pal 10. This image measures
3\farcm0$\times$3\farcm0. North is up and east is to the left. The numbering scheme
is the same as that of Table 1.
\label{fig6}}

\plotone{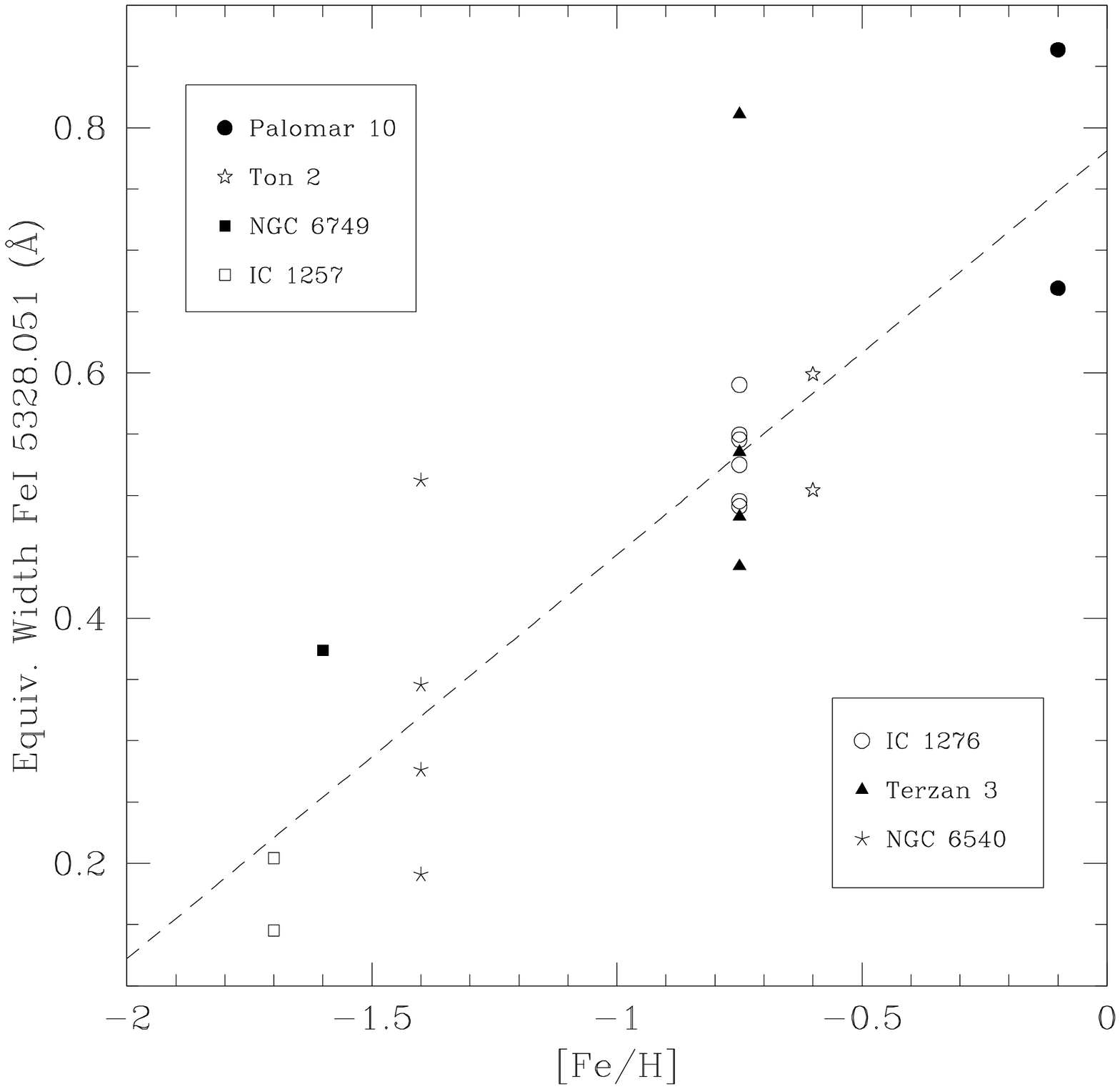}

\figcaption[gcskin.07.ps]{Equivalent width of the FeI 5328.051 \AA\ line measured from our
HIRES spectra plotted against published [Fe/H] for red giant stars in Pal 10, Ton 2, 
IC 1257 and NGC 6749. The dashed line shows the least-squares fit to the two quantities.
Based on this relation, we estimate [Fe/H] $\sim -0.75\pm0.30$ for Terzan 3, 
[Fe/H] $\sim -0.75\pm0.15$ for IC 1276,
and suggest a downward revision of the adopted metallicity of NGC 6540 from
[Fe/H] = $-1.0$ to $-1.40\pm0.25$.
\label{fig7}}

\plotone{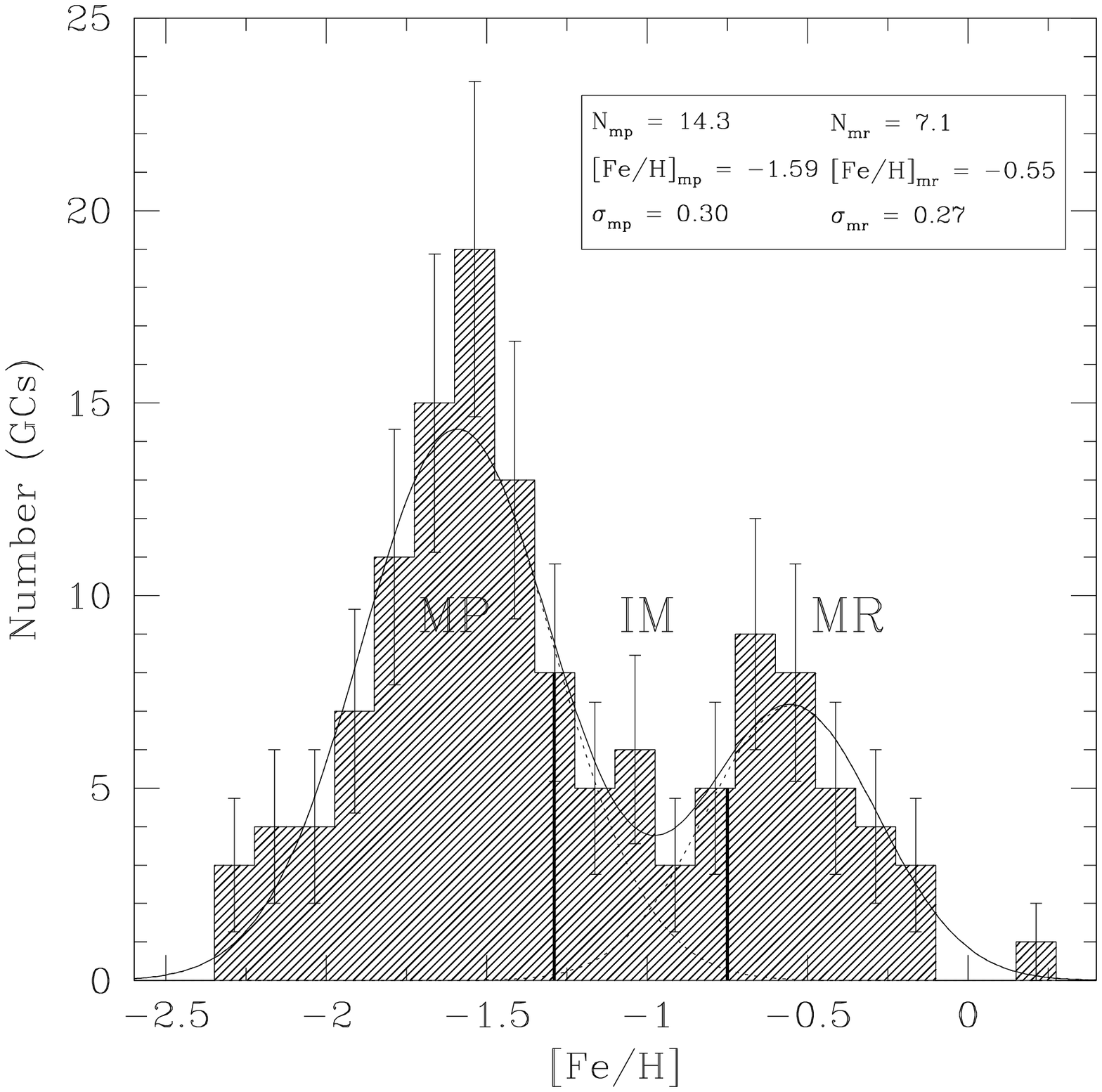}

\figcaption[gcskin.08.ps]{Histogram of 133 Galactic GCs having measured
metallicities. The two-component Gaussian which best fits the observed distribution is 
shown as the solid line. The separate metal-rich and
metal-poor components are indicated by the dotted lines. The heavy vertical lines at
[Fe/H] = $-1.29$ and $-0.75$ show the boundaries used to define the samples of metal-rich (MR),
metal-poor (MP) and intermediate metallicity (IM) clusters.
\label{fig8}}

\plotone{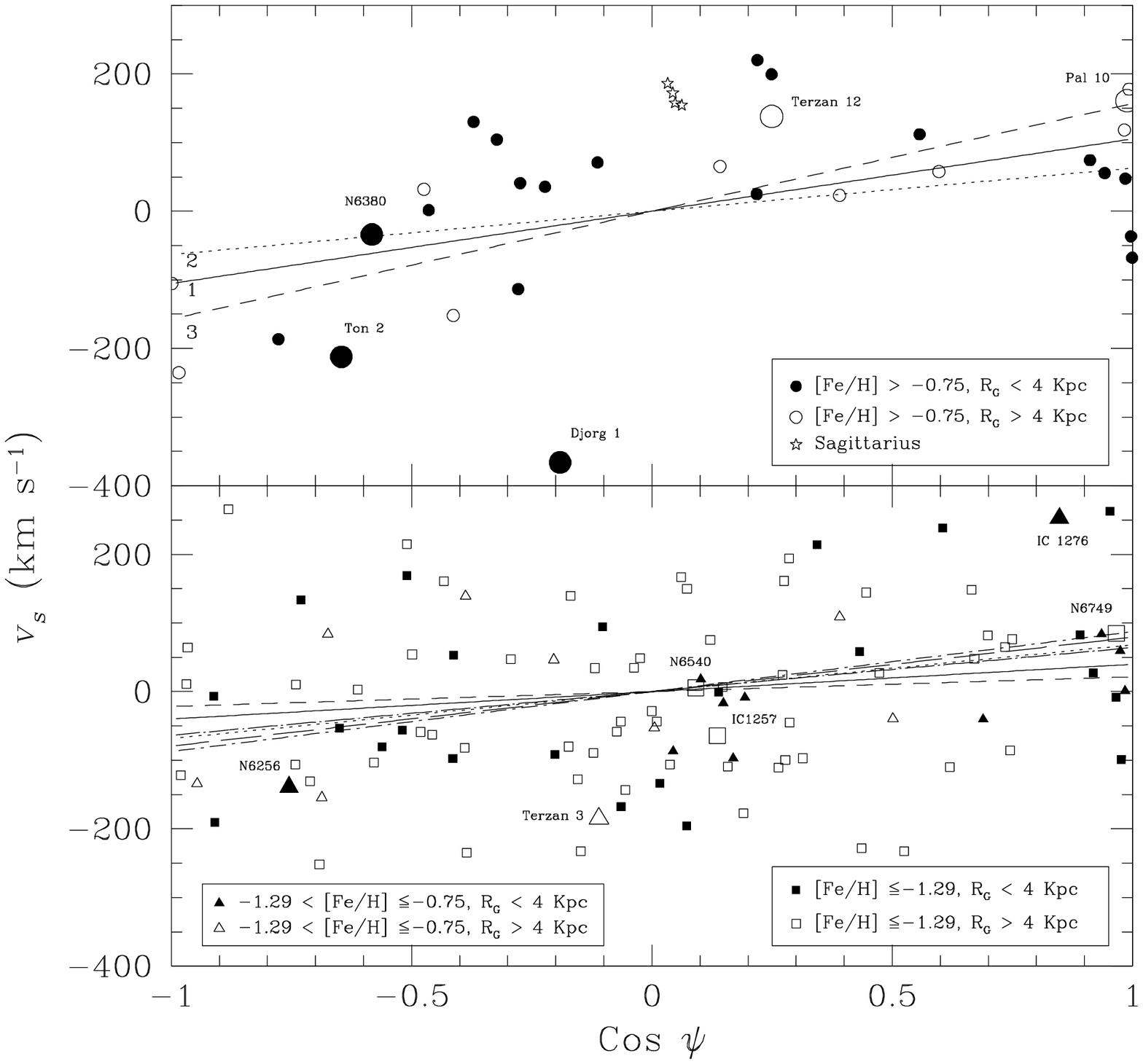}

\figcaption[gcskin.09.ps]{(Upper panel)
Plot of $v_s$ versus cos$\psi$ for metal-rich GCs. Here $\psi$ is the
angle between the rotational-velocity vector of the cluster and the direction to the observer,
and $v_s$ is the velocity with respect to a stationary observer at the location of the Sun.
Objects having a rotation velocity which is constant with Galactocentric radius 
should fall along a straight line in this diagram.
The symbols are described in the keys to the figure. Small symbols refer to GCs
taken from the catalog of Harris (1996); large symbols indicate clusters with
radial velocities reported in this paper. The four probable or suspected members 
of the Sagitarrius dwarf galaxy have been omitted from kinematic analysis of the 
Galactic GC system. The straight lines show the best-fit rotation solutions of the
form of equation 3; each relation is numbered according scheme given in Table 4.
(Lower Panel) Same as above, except for metal-poor GCs (squares) and intermediate metallicity GCs (triangles).
\label{fig9}}

\plotone{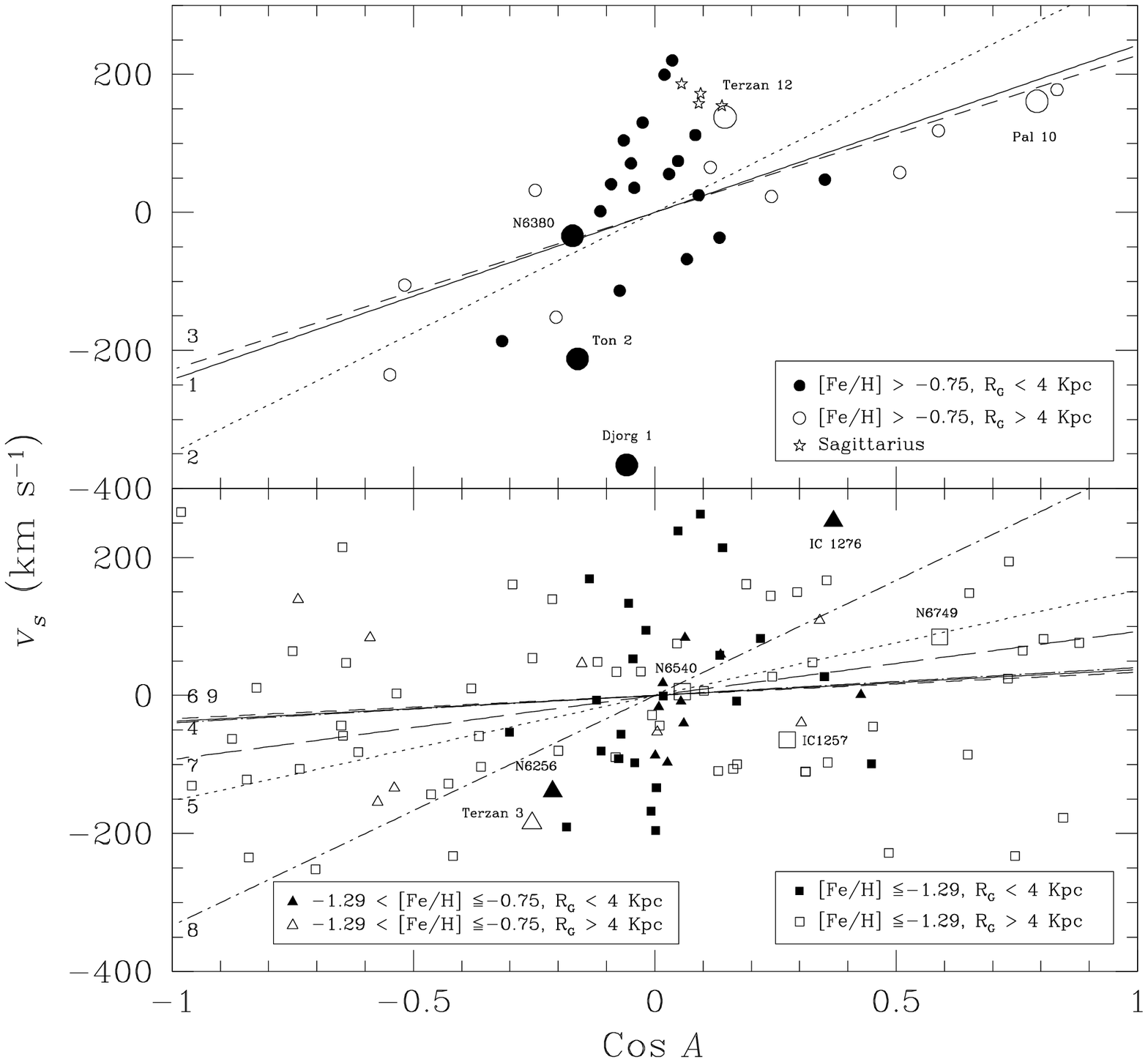}

\figcaption[gcskin.10.ps]{(Upper Panel) Plot of $v_s$ versus cos$A$ for Galactic GCs, 
where $A$ is the angle between the apex
of the Local Standard of Rest and the cluster's position on the sky. The symbols are the
same as in Figure 9. Objects exhibiting solid-body rotation should fall along a straight
line in this diagram.
(Lower Panel) Same as above, except for 
metal-poor GCs (squares) and intermediate metallicity GCs (triangles).
\label{fig10}}

\plotone{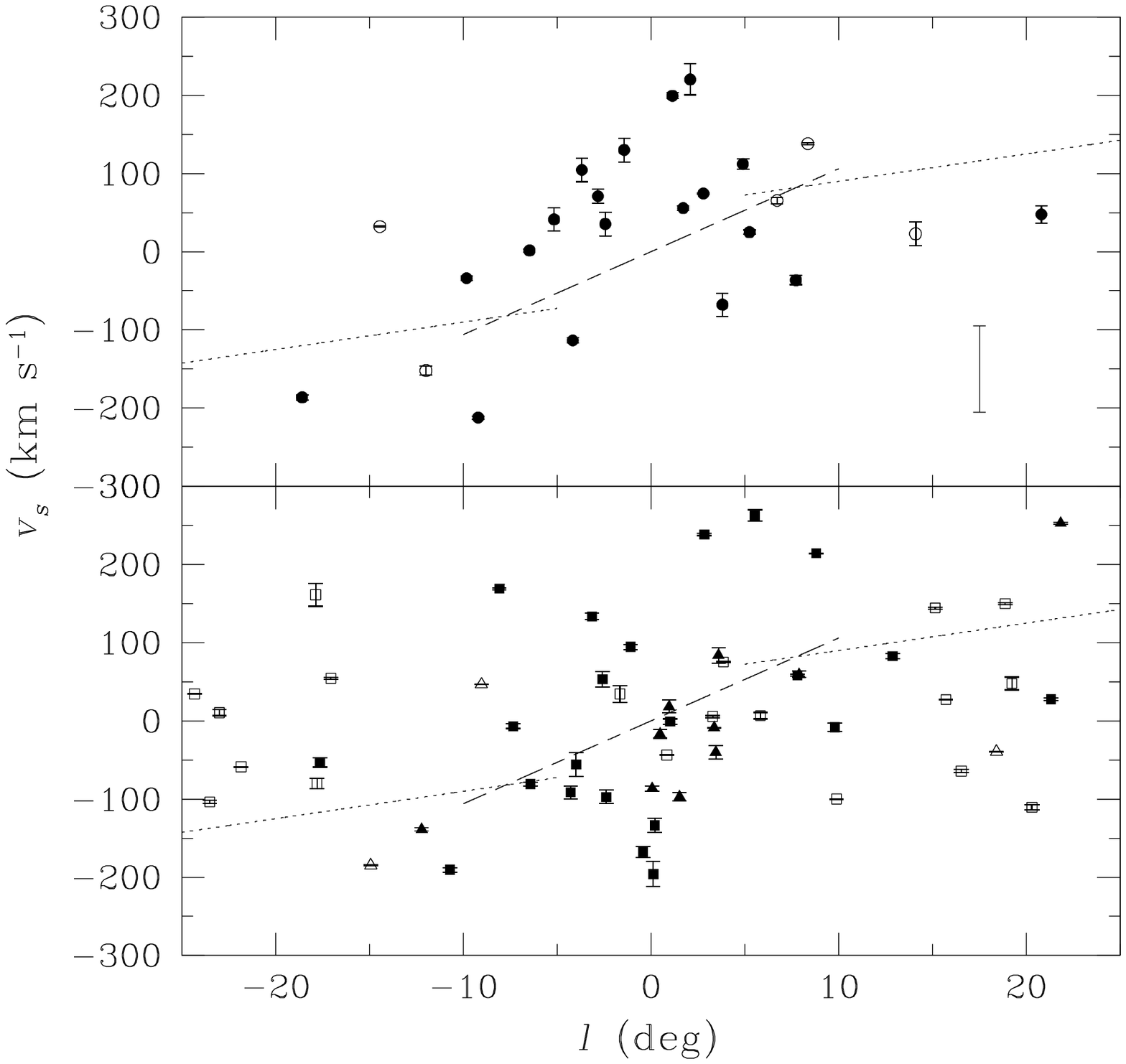}

\figcaption[gcskin.11.ps]{(Upper Panel) $v_s$ plotted against Galactic longitude for 
metal-rich GCs. Open circles indicate GCs having $R_G \ge 4$ kpc; filled circles
depict GCs having $R_G < 4$ kpc. The dashed line shows the rotation law for inner bulge stars 
reported by Tiede \& Terndrup (1997). The dotted lines indicate the outer bulge rotation
of $\sim$ 25 km s$^{-1}$ kpc$^{-1}$ found by Ibata \& Gilmore (1995), shifted by $\pm$ 55 
km s$^{-1}$ to match the inner relation of Tiede \& Terndrup (1997).
The errorbar shows the typical velocity dispersion of bulge objects (Kent 1992).
(Lower Panel) $v_s$ plotted against Galactic longitude for metal-poor GCs (squares)
and intermediate metallicity GCs (triangles). Open symbols refer to GCs
having $R_G \ge 4$ kpc; filled symbols indicate GCs having $R_G < 4$ kpc.
\label{fig11}}

\plotone{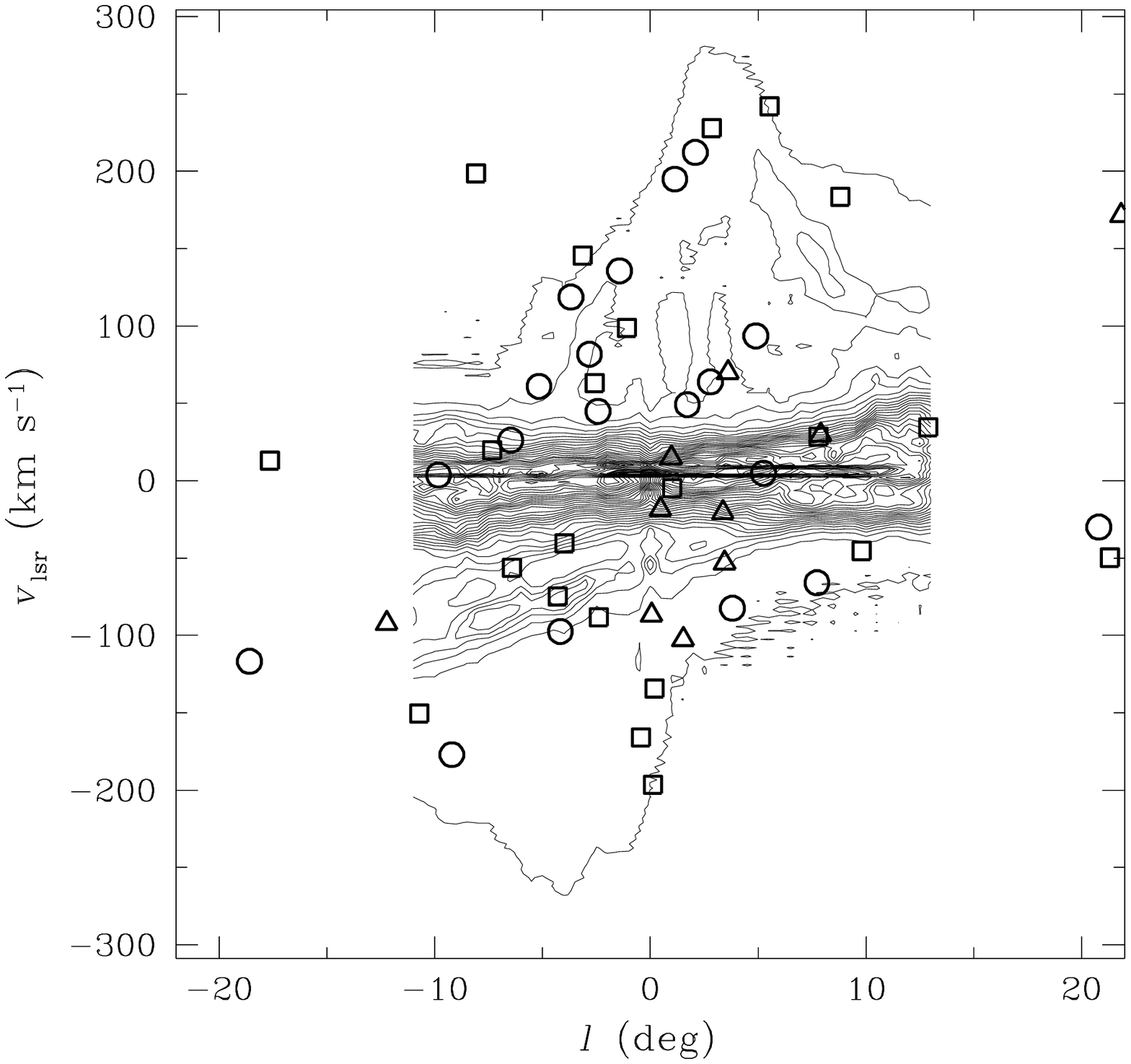}

\figcaption[gcskin.12.ps]{Longitude-velocity diagram in the direction of the Galactic center. 
The contours indicate the HI data of Burton \& Liszt (1983), summed over the interval 
$|b| \le 1.5^{\circ}$.  Metal-rich GCs are indicated by circles, metal-poor GCs by
squares and intermediate-metallicity GCs by triangles. 
With the exception of Djorg 1 (see text), all GCs within 4 kpc of the Galactic center are
shown on this figure.
The metal-poor cluster at $l \simeq -8^{\circ}$ and $v_{\rm lsr} \simeq 200$ km s$^{-1}$ is NGC 6144.
\label{fig12}}

\plotone{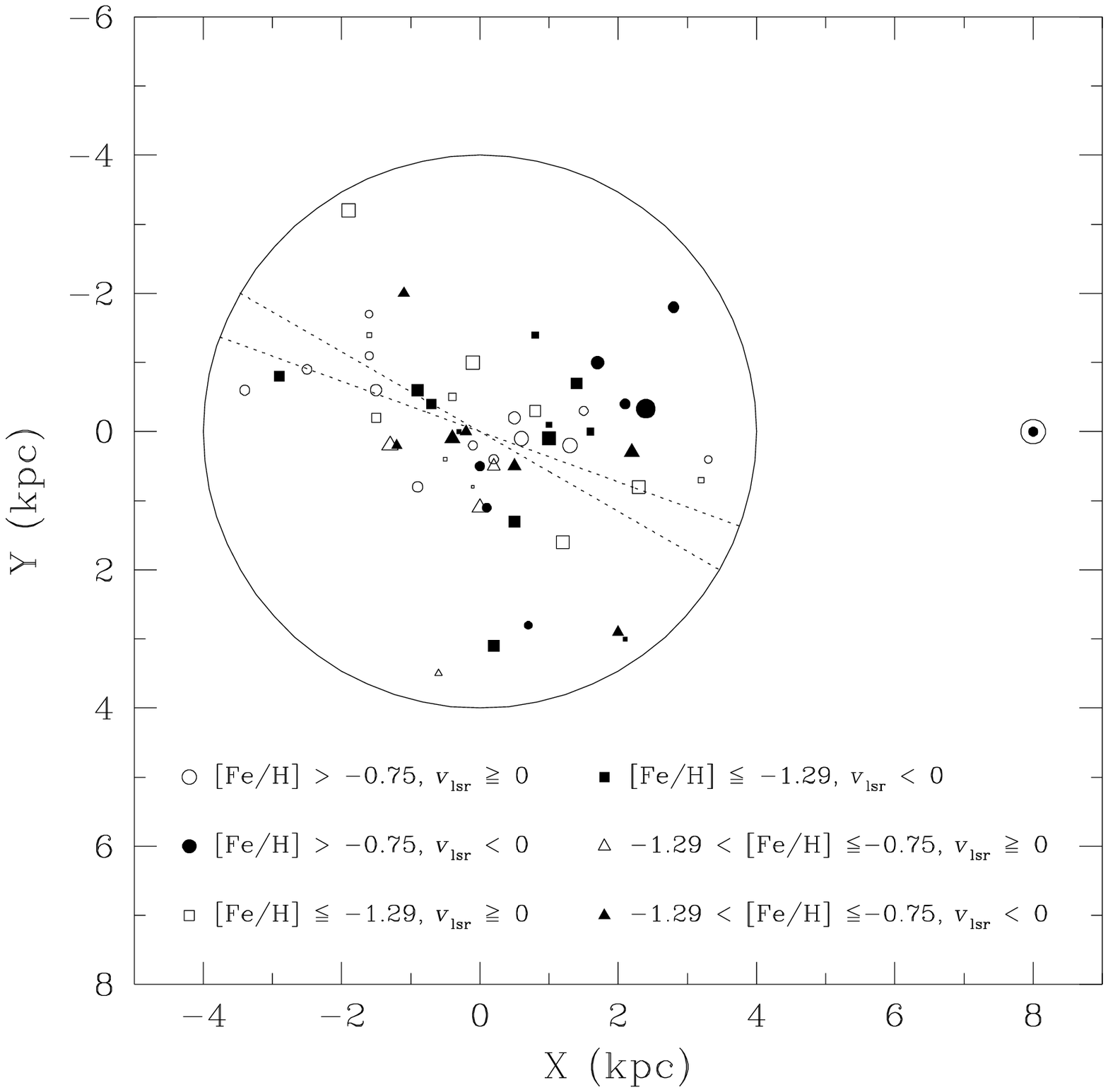}

\figcaption[gcskin.13.ps]{Distribution of all GCs within 4 kpc of the Galactic center having measured
radial velocities. Metal-rich GCs are indicated by circles, metal-poor GCs by squares and 
intermediate-metallicity GCs by triangles. Open and filled symbols indicate GCs having 
$v_{\rm lsr} \ge 0$ and $v_{\rm lsr} < 0$, respectively. The size of the symbol is proportional
to the absolute value of the observed velocity.
The position of the sun is indicated
at (8,0). The large circle with radius 4 kpc shows the approximate co-rotation radius of the 
Galactic bar (Fux 1997), while the two dotted lines indicate the limits on the orientation of the Galactic 
bar based on observations of red clump giants (Stanek et al. 1997).
\label{fig13}}

% Option 2.  The figure captions are printed on a caption page(s) as in 
% option 1.  The figures available as EPS files are then printed at the
% end of the document, one figure per page, using the \plotone command.
% If you wish to process this option then simply comment out the \end{document}
% just above these five lines. 
 
\end{document}